\documentclass[conference]{IEEEtran}
\IEEEoverridecommandlockouts
\usepackage{cite}
\usepackage{amsmath,amssymb,amsfonts}
\usepackage{algorithmic}
\usepackage{graphicx}
\usepackage{textcomp}
\usepackage{braket}
\usepackage{url}
\usepackage[dvipsnames]{xcolor}
\usepackage{svg}
\usepackage{bm}
\usepackage{hyperref}
\usepackage{subcaption}
\usepackage{tikz}
\usetikzlibrary{positioning,decorations.markings,decorations.pathreplacing,calc,shapes.geometric,arrows,fit}
\makeatletter
\pgfdeclareshape{document}{
\inheritsavedanchors[from=rectangle] %
\inheritanchorborder[from=rectangle]
\inheritanchor[from=rectangle]{center}
\inheritanchor[from=rectangle]{north}
\inheritanchor[from=rectangle]{south}
\inheritanchor[from=rectangle]{west}
\inheritanchor[from=rectangle]{east}
\backgroundpath{%
\southwest \pgf@xa=\pgf@x \pgf@ya=\pgf@y
\northeast \pgf@xb=\pgf@x \pgf@yb=\pgf@y
\pgf@xc=\pgf@xb \advance\pgf@xc by-10pt %
\pgf@yc=\pgf@yb \advance\pgf@yc by-10pt
\pgfpathmoveto{\pgfpoint{\pgf@xa}{\pgf@ya}}
\pgfpathlineto{\pgfpoint{\pgf@xa}{\pgf@yb}}
\pgfpathlineto{\pgfpoint{\pgf@xc}{\pgf@yb}}
\pgfpathlineto{\pgfpoint{\pgf@xb}{\pgf@yc}}
\pgfpathlineto{\pgfpoint{\pgf@xb}{\pgf@ya}}
\pgfpathclose
\pgfpathmoveto{\pgfpoint{\pgf@xc}{\pgf@yb}}
\pgfpathlineto{\pgfpoint{\pgf@xc}{\pgf@yc}}
\pgfpathlineto{\pgfpoint{\pgf@xb}{\pgf@yc}}
\pgfpathlineto{\pgfpoint{\pgf@xc}{\pgf@yc}}
}
}
\makeatother

\usepackage{listings}
\definecolor{mygreen}{rgb}{0,0.6,0}
\definecolor{mygray}{rgb}{0.5,0.5,0.5}
\definecolor{mymauve}{rgb}{0.58,0,0.82}
\lstdefinestyle{floatingcode}{
  float,
  floatplacement=tbp
}

\usepackage[normalem]{ulem}
\definecolor{Burgundy}{RGB}{144,0,32}

\hypersetup{
    colorlinks=true,
    linkcolor=blue,
    filecolor=magenta,      
    urlcolor=blue,
    }

\newcommand{\qaoakit}{\texttt{QAOAKit}}
\newcommand{\qaoakitdata}{\texttt{QAOAKit\_data}}

\def\BibTeX{{\rm B\kern-.05em{\sc i\kern-.025em b}\kern-.08em
    T\kern-.1667em\lower.7ex\hbox{E}\kern-.125emX}}
\begin{document}

\title{\qaoakit{}: A Toolkit for Reproducible Study, Application, and Verification of the QAOA\\ 
\thanks{RS was supported by the U.S.\ Department of Energy (DOE), Office of Science, Office of Advanced Scientific Computing Research AIDE-QC and FAR-QC projects and by the Argonne LDRD program under contract number DE-AC02-06CH11357. JW was supported by the Defense Advanced Research Projects Agency (DARPA) under Contract No. HR001120C0068, and NSF STAQ project (PHY1818914). PCL was supported by the Defense Advanced Research Project Agency through DOE project 1868-Z361-20.
KM was supported by the National Science Foundation Graduate Research Fellowship Program under Grant No. DGE-1746045.
}
}

\author{\IEEEauthorblockN{Ruslan Shaydulin\IEEEauthorrefmark{1}, Kunal Marwaha\IEEEauthorrefmark{2}\IEEEauthorrefmark{3}, Jonathan Wurtz\IEEEauthorrefmark{4} and Phillip C. Lotshaw\IEEEauthorrefmark{5}}\vspace{0.05in}
\IEEEauthorblockA{\IEEEauthorrefmark{1}Mathematics and Computer Science Division, Argonne National Laboratory, Lemont, IL 60439 USA}
\IEEEauthorblockA{\IEEEauthorrefmark{2}Department of Computer Science, University of Chicago, Chicago, IL 60637 USA}
\IEEEauthorblockA{\IEEEauthorrefmark{3}Berkeley Center for Quantum Information and Computation, University of California, Berkeley, CA 94720 USA} 
\IEEEauthorblockA{\IEEEauthorrefmark{4}Department of Physics and Astronomy, Tufts University, Medford, MA 02155 USA}
\IEEEauthorblockA{\IEEEauthorrefmark{5}Quantum Computational Sciences Group, Oak Ridge National Laboratory, Oak Ridge, TN 37830 USA}
Email: \IEEEauthorrefmark{1}ruslan@anl.gov, \IEEEauthorrefmark{2}kmarw@uchicago.edu, \IEEEauthorrefmark{4}jonathan.wurtz@tufts.edu \IEEEauthorrefmark{5}lotshawpc@ornl.gov}

\maketitle

\begin{abstract}
Understanding the best known parameters, performance, and systematic behavior of the Quantum Approximate Optimization Algorithm (QAOA) remain open research questions, even as the algorithm gains popularity. We introduce \qaoakit{}, a Python toolkit for the QAOA built for exploratory research. \qaoakit{} is a unified repository of preoptimized QAOA parameters and circuit generators for common quantum simulation frameworks.
We combine, standardize, and cross-validate previously known parameters for the MaxCut problem, and incorporate this into \qaoakit{}.
We also build conversion tools to use these parameters as inputs in several quantum simulation frameworks that can be used to reproduce, compare, and extend known results from various sources in the literature. We describe \qaoakit{} and provide examples of how it can be used to reproduce research results and tackle open problems in quantum optimization.

\end{abstract}


\tikzstyle{storage} = [cylinder, draw,minimum height=0.5cm,minimum width=2cm]
\tikzstyle{arrow} = [thick,->,>=stealth]
\tikzstyle{doc}=[%
draw,
align=center,
color=black,
shape=document,
minimum width=20mm,
minimum height=2cm,
shape=document,
inner sep=2ex,
]
\lstset{basicstyle=\small\ttfamily,columns=fullflexible}

\section{Introduction}
Optimization is considered one of the most promising applications for useful quantum computing. A popular candidate algorithm is the Quantum Approximate Optimization Algorithm (QAOA)~\cite{Hogg2000,farhi2014quantum}. The QAOA is a hybrid quantum-classical algorithm that combines a parameterized quantum evolution with a classical procedure for determining good parameters. The measurement outcomes of the quantum evolution with optimized parameters correspond to the solutions of the original classical optimization problem. 

The choice of parameters (commonly referred to as angles) for the QAOA is of central importance to algorithm performance and has been considered extensively in the literature; see, for example,~\cite{Shaydulin2019MultistartDOI,2108.05288,zhou2018qaoaperformance}. Despite the abundance of results, reproducibility is challenging; researchers use different simulation implementations with no standardized parameterization. This makes it hard to leverage existing datasets of optimized parameters to build on the previous state-of-the-art. Data reuse is especially useful because recent results show that QAOA parameters can be transferred across graphs. These results suggest that the QAOA can be executed completely optimization-free by using a dataset of preoptimized parameters~\cite{brandao2018fixed-rs,Shaydulin2019MultistartDOI,2106.07531,zhou2018qaoaperformance} or a pretrained machine learning model~\cite{khairy2019learning,wilson2019optimizing,verdon2019learning}. There is a need for a standardized approach to share such parameters.

In this paper we introduce \qaoakit{},  a repository of standardized, preoptimized QAOA parameters and reference QAOA circuit implementations. We design \qaoakit{} with flexibility and ease of use in mind and seed it with the largest dataset of preoptimized parameters for the QAOA applied to the MaxCut problem ever assembled in one place. We provide examples of how \qaoakit{} can be leveraged by practitioners to obtain high-quality initial guesses for parameter optimizations and by researchers to reproduce results in the literature and tackle open problems in quantum optimization. We make \qaoakit{} and all corresponding data available on GitHub and encourage contributions: \texttt{\url{https://github.com/QAOAKit/QAOAKit}}.  \qaoakit{} is available on PyPI and easy to install with the following two lines:

\begin{lstlisting}[language=bash]
pip install qaoakit
python -m QAOAKit.build_tables
\end{lstlisting}

\begin{figure*}\centering
\begin{tikzpicture}[node distance=0.4cm]
\node (dataset1) [storage] {Raw dataset};
\node (dataset2) [storage, below=of dataset1] {Raw dataset};
\node (compiled_dataset1) [storage, left=2cm of dataset1] {Compiled dataset};
\node (compiled_dataset2) [storage, below=of compiled_dataset1] {Compiled dataset};
\node (high_api) [doc, left=2cm of compiled_dataset1] {High-level API\\ Conversion tools \\ QAOA circuits};
\node (low_api) [doc, below=of high_api] {Low-level API\\ (easy access \\to full datasets)};
\node[draw, thick, dotted, rounded corners, inner xsep=1em, inner ysep=1em, fit=(high_api) (compiled_dataset2)] (qaoakitdata) {};
\node[fill=white] at (qaoakitdata.north) {\qaoakit{}};
\node[draw, thick, dotted, rounded corners, inner xsep=1em, inner ysep=1em, fit=(dataset1) (dataset2)] (qaoakit) {};
\node[fill=white] at (qaoakit.north) {\texttt{QAOAKit\_data}};
\path (dataset1) -- node[auto=false]{\ldots} (dataset2);
\draw [arrow] (dataset1) -- node [fill=white,above] {Compiled} node [fill=white,below] {at install} (compiled_dataset1);
\draw [arrow] (dataset2) -- node [fill=white,above] {Compiled} node [fill=white,below] {at install} (compiled_dataset2);
\path (compiled_dataset1) -- node[auto=false]{\ldots} (compiled_dataset2);
\draw [arrow] (compiled_dataset1) -- (high_api);
\draw [arrow] (compiled_dataset1) -- (low_api);
\draw [arrow] (compiled_dataset2) -- (high_api);
\draw [arrow] (compiled_dataset2) -- (low_api);
\end{tikzpicture}
\caption{High-level overview of \qaoakit{}, containing high- and low-level APIs (left) compiled from a variety of datasets (middle and right). The data in the separate repository \qaoakitdata{}~\cite{qaoakitdata} is available in human-readable files (CSV or JSON). At \qaoakit{} install time the datasets are compiled into binaries to improve runtime and augmented with additional data for easier use. For example, we augment MaxCut data with graph isomorphism certificates to accelerate retrievals of dataset entries corresponding to a given graph. A high-level API provides commonly used functions such as retrieving optimal parameters for a given graph and converting them to the formats used in common frameworks (QTensor and Qiskit are currently available). A low-level API exposes complete datasets to researchers for in-depth analysis.}
\label{fig:qaoakit_overview}
\end{figure*}
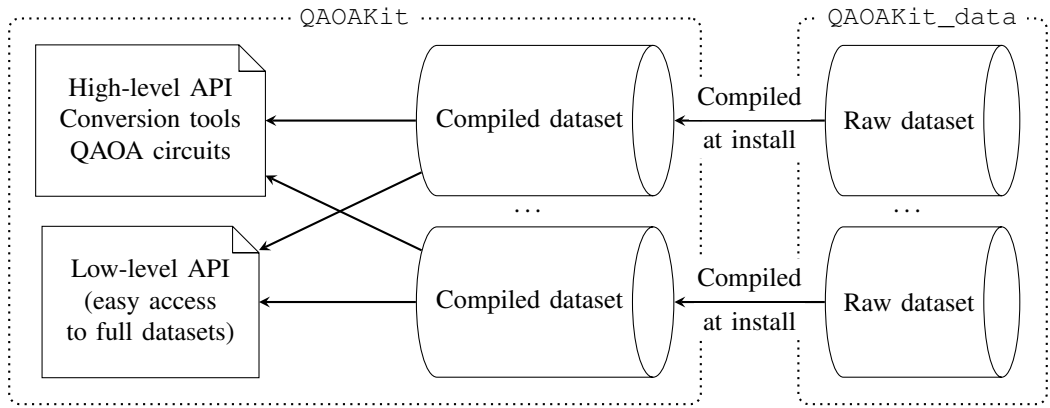

\section{\qaoakit{} overview and datasets}

\qaoakit{} is designed as a lightweight and easy-to-use repository for optimal QAOA parameters. An overview of the \qaoakit{} structure is provided in Figure~\ref{fig:qaoakit_overview}. \qaoakit{} is targeted primarily at researchers. We seed \qaoakit{} with the following two datasets. %

First, we include the dataset from \cite{lotshaw2021empirical}, which provides an exhaustive set of all connected nonisomorphic graphs of sizes $n\leq 9$ \cite{GraphFiles}. For each graph, the dataset includes QAOA cost values and angles for each graph at $p \leq 3$. These were computed for each graph by using the best result from 50 ($p=1$), 100 ($p=2$), or 1,000 $(p=3)$ calls to the Broyden--Fletcher--Goldfarb--Shanno algorithm with random angle seeds.  The results for $p=1$ were verified as optimal solutions against exact approaches. 

Second, we include the dataset from \cite{2107.00677}, which provides an exhaustive set of all connected nonisomorphic 3-regular graphs of sizes $n\leq 16$. There are 4,681 such unique graphs \cite{Meringer1999}. For each graph, the dataset includes QAOA cost values and all degenerate maximal parameters for $p=1$ and $2$. These were found through a multistart grid search gradient ascent routine, which returns all degenerate maxima. Additionally, the dataset includes the fixed angles of \cite{2107.00677} for all $p\leq 11$ and the QAOA cost values evaluated at these angles. 
The approximation ratios attained by the QAOA with these angles have non-trivial lower bounds under the fixed angle conjecture \cite{Wurtz2020bounds}. It was
numerically observed in \cite{Wurtz2020bounds} that these fixed angles are close to optimal and so can be a good proxy for QAOA performance at intermediate $p$.

Combining multiple independently obtained datasets allows us to perform additional validation by cross-referencing the overlapping parts of the datasets (3-regular graphs with up to $9$ nodes). \qaoakit{} makes it easy to retrieve a set of optimal angles and run the corresponding circuit in a preferred simulation framework or real quantum device, as shown in Listing~\ref{lst:get_opt_params}. The parameters are retrieved by computing a graph isomorphism certificate for the input graph using \texttt{nauty}~\cite{nautytracesmanual} and using it as a hash to retrieve the angles from the database. If for a given instance no optimal angles are available in any of the datasets, the closest fixed angles are returned~\cite{Wurtz2020bounds}. We anticipate the number of included datasets to grow in the future and to extend beyond the MaxCut problem.

\begin{lstlisting}[language=Python, caption={A code example using \qaoakit{} to generate and run a QAOA circuit with optimal parameters.}, label={lst:get_opt_params}, style=floatingcode]
# build graph
import networkx as nx 
G = nx.star_graph(5)
# grab optimal angles
from QAOAKit import opt_angles_for_graph, angles_to_qaoa_format
p = 3
angles = angles_to_qaoa_format(
    opt_angles_for_graph(G,p))
# build circuit and print measurement outcomes
from QAOAKit.qaoa import get_maxcut_qaoa_circuit
qc = get_maxcut_qaoa_circuit(
    G, angles['beta'], angles['gamma'])
qc.measure_all()
# run circuit
from qiskit.providers.aer import AerSimulator
backend = AerSimulator()
print(backend.run(qc).result().get_counts())
\end{lstlisting}

\section{Applications}

We illustrate the value of \qaoakit{} by providing  examples of how it can be used to verify and extend findings from the QAOA literature. A simple code example demonstrating the use of  \qaoakit{} to access optimal angles and circuits is shown in Listing \ref{lst:get_opt_params}. Beyond circuit generation, we demonstrate several examples, which can be accessed in \url{https://github.com/QAOAKit/QAOAKit/blob/master/examples/}.

In Secs.~\ref{sec:concentration_symm} and  ~\ref{sec:param_symmetries} we showcase how \qaoakit{} can be used to study the symmetries in QAOA parameters. In Sec.~\ref{sec:trasfer} we show how this concentration can be used to obtain near-optimal parameters for instances not present in the dataset. This provides a practical application of \qaoakit{} for reducing the cost of running the QAOA. Moreover, we show how the the data in \qaoakit{} illuminates the performance of the QAOA and its relationship with different properties of the graph instances such as density (Sec.~\ref{sec:performance}) and symmetry (Sec.~\ref{sec:features}). In Sec.~\ref{sec:classical} we show how performance of the QAOA averaged over the large-scale available datasets can be compared with state-of-the-art classical algorithms implemented in \qaoakit{} and elsewhere.

\begin{figure}
    \centering
    \includegraphics[width=\linewidth]{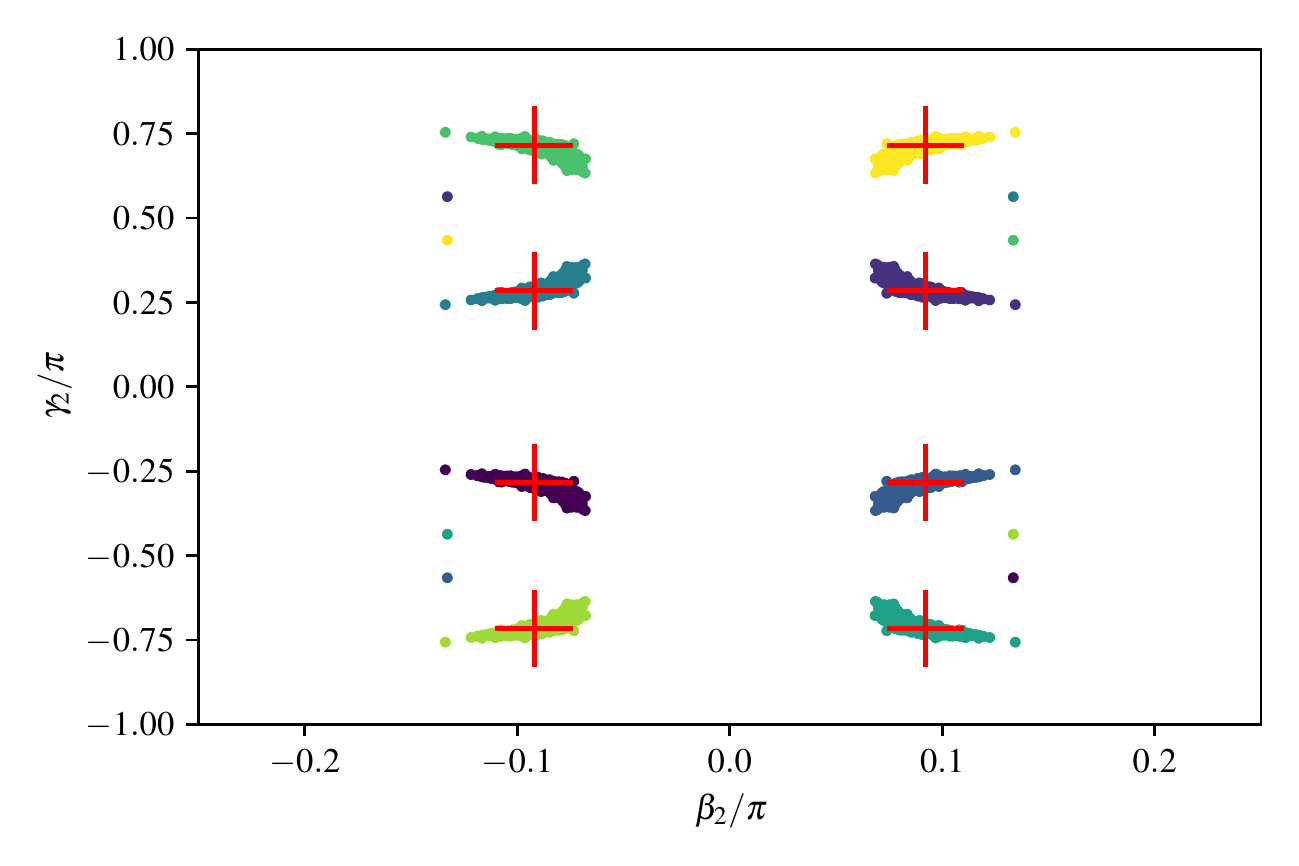}
    \caption{Degenerate optima for 3 regular graphs, clustered into the 8 symmetry sectors of $p=2$. Red ``+'' marks indicate the center of each cluster, while colors indicate the cluster assignment of each angle. One can  clearly  see that even within each symmetry sector, the values concentrate around the mean value; for $p=1$ the RMS distance is $0.00546$, while for $p=2$ the RMS distance is $0.01903$.}
    \label{fig:concentration}
\end{figure}

\subsection{Concentration within symmetry sectors} \label{sec:concentration_symm}

A well-documented property of the QAOA is that of concentration of both optimal parameters and energy landscapes in general~\cite{brandao2018fixed-rs}.  It has been observed that within an ensemble of graphs, the optimal value of each graph is close to the optimal angles of every other graph. To demonstrate \qaoakit{}, we illustrate this feature by analyzing the dataset of all 3-regular graphs with $\leq 16$ vertices. Observing concentration of parameters is complicated by the fact that each set of optimal angles is degenerate with an infinite number of others, under the $Z_2$ symmetry of $\beta\mapsto \beta\pm \pi/2$ and the $U(1)$ symmetry of $\gamma\mapsto \gamma \pm 2\pi$.
Because of these symmetries, the distance between each angle is  well defined only within a particular choice of symmetry sector. The function \texttt{QAOAKit.get\_3\_reg\_dataset\_table} returns a pandas DataFrame containing the angles within the symmetry sector $\gamma\in[-\pi,\pi]$, $\beta\in [-\pi/4,\pi/4]$. However, there are additional symmetries within the $Z_2$ and $U(1)$ sector. For the 3 regular datasets and $p=1$, we observe that there are 4 degenerate optima; and for $p=2$, there are 8 degenerate optima. Included in \qaoakit{} is a low-level API that provides, for each graph, every optimal angle within the $Z_2$ and $U(1)$ symmetric region of parameter space. With this dataset, we may compute concentration within each symmetry sector.

	\begin{figure}
	\centering
	\begin{subfigure}{.24\textwidth}
  \centering
  \includegraphics[width=.95\linewidth]{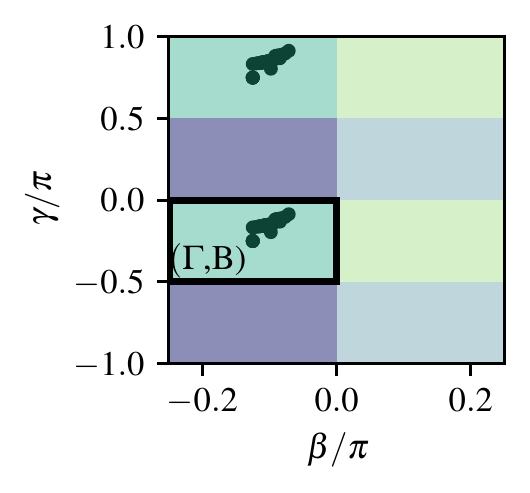}
  \caption{even}
  \label{fig:edgesub1}
\end{subfigure}%
\begin{subfigure}{.24\textwidth}
  \centering
  \includegraphics[width=.95\linewidth]{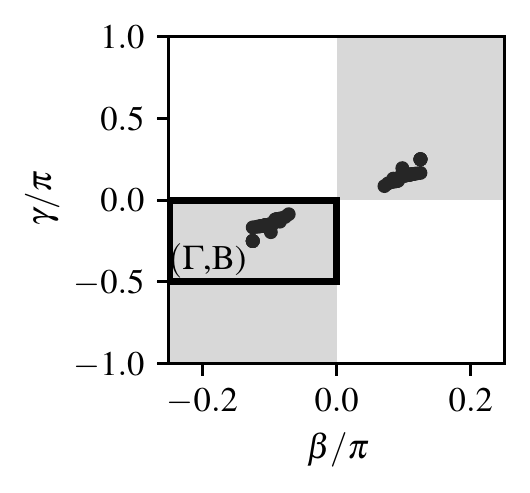}  \caption{time-reversal}
  \label{fig:edgesub2}
\end{subfigure}
\begin{subfigure}{.24\textwidth}
  \centering
  \includegraphics[width=.95\linewidth]{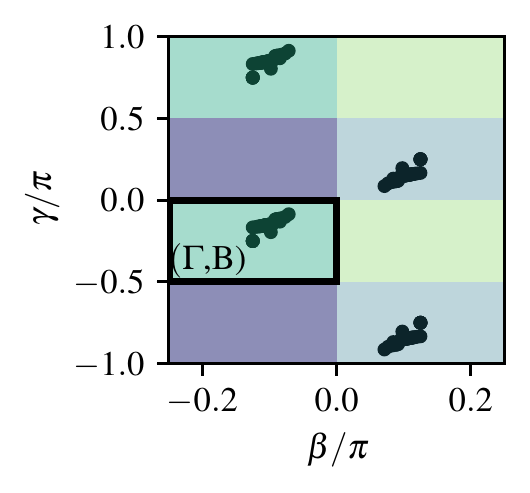}
  \caption{even \& time-reversal}
  \label{fig:edgesub3}
\end{subfigure}
\begin{subfigure}{.24\textwidth}
  \centering
  \includegraphics[width=.95\linewidth]{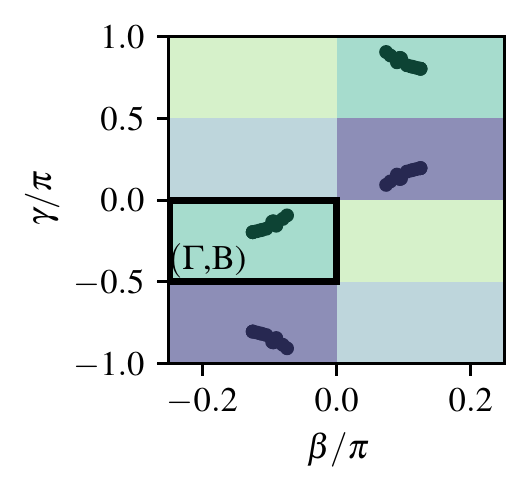}
  \caption{odd \& time-reversal}
  \label{fig:edgesub4}
\end{subfigure}
\caption{Symmetry-related optimized angles for regular graphs with even-degree symmetry (a), time-reversal symmetry (b), even and time-reversal (c), and odd and time-reversal (d). Colored blocks indicate symmetry-related sections of parameter space.}
\label{fig:symmetry}
\end{figure}

\begin{table}
\center
\begin{tabular}{ |c|c| }%
 \hline
 Graph & Angle Symmetry \\
 \hline
 All & $\bm \gamma, \bm \beta \to -\bm \gamma, -\bm \beta$ \\
 \hline
 Even & $\gamma_l \to \gamma_l \pm \pi$ \\
 \hline
  Odd & $\gamma_l \to \gamma_l \pm \pi$, $\beta_q \to -\beta_q\ \forall q \geq l$ \\
 \hline
\end{tabular}
\caption{Angle symmetries in MaxCut QAOA \cite{lotshaw2021empirical}. ``Even" and ``odd" refer to graphs where each vertex degree is even or odd.}
\label{symmetrytable}
\end{table}

The problem of assigning angles to symmetry sectors may be solved using a clustering algorithm, a common unsupervised machine learning task. In the example script \href{https://github.com/QAOAKit/QAOAKit/blob/master/examples/degenerate_optima_in_angle_space.py}{degenerate\_optima\_in\_angle\_space.py} we use an implementation of the well-known K-means clustering algorithm~\cite{scikit-learn} to find average angles within each symmetry sector. First, the script compiles all degenerate angles for each of the 4,681 graphs into a 2d array: one row for each graph and degenerate optima and one column for each parameter $\gamma,\beta$. This is a $4681*4\,\times\,2$ array for $p=1$, and a $4681*8\,\times\,4$ array for $p=2$. Then, the script clusters these degenerate angles into 4 or 8 clusters, respectively, using the K-means algorithm.

Some results of the clustering are shown in Figure~\ref{fig:concentration}, in the $(\beta_2,\gamma_2)$ plane of $p=2$. The red ``+'' marks indicate the center of each of the 8 symmetry sectors, while colors indicate the cluster assignment of each degenerate angle of every graph in the ensemble. Crucially, we observe that within each sector the angles strongly cluster about the center of each cluster, consistent with the phenomenon of concentration. A simple measure of this concentration is the root mean square (RMS) distance between each point and the center, measured as $\Delta = \sqrt{\sum_i(\Theta_i-\Theta_c)^2/N}$, which K-means clustering minimizes. For $p=1$ (not shown because of the strength of concentration) the RMS distance is $0.00546$, while for $p=2$ the RMS distance is $0.01903$. These distances are small in units of $1$, indicating that the values concentrate. Choosing the symmetry sector such that the total angular magnitude is smallest and all angles have positive values, the cluster centers are

\begin{equation}
    \begin{tabular}{clc}
         $(\beta_1,\gamma_1)$ &$=$&$(\,0.120,\;0.191\,)\pi,$ \\
         $(\beta_1,\beta_2,\gamma_1,\gamma_2)$&$=$&$(\,0.169,\,0.092,\,0.159,\,0.284\,)\pi.$
    \end{tabular}\nonumber
\end{equation}

\subsection{Even, odd, and time-reversal symmetries} \label{sec:param_symmetries}

Using \qaoakit{} one can easily  generate and visualize the symmetries of angles and the variantion in symmetry depending on the graph structure. To do so, we begin with optimized component angles $-\pi \leq \gamma_l \leq \pi$ and $-\pi/4 \leq \beta_l \leq \pi/4$ from the dataset of \cite{lotshaw2021empirical}, which can be loaded from \qaoakit{} as a pandas DataFrame using the function \texttt{get\_full\_qaoa\_dataset\_table}. We map these to symmetry-related angles following the symmetries of Table \ref{symmetrytable}. The time-reversal symmetry holds for all graphs, so there are two degenerate sets of angles for all graphs.  There are also distinct symmetries for graphs where each vertex degree is even or each degree is odd, giving four sets of angles for these graphs at $p=1$ (see~\cite{lotshaw2021empirical} for details). With \qaoakit{} we can easily visualize the different sets of angles for graphs under these varying symmetries to understand their similarities and differences. 

In the example script \href{https://github.com/QAOAKit/QAOAKit/blob/master/examples/even_odd_angle_symmetry.ipynb}{even\_odd\_angle\_symmetry.ipynb} we use \qaoakit{} and \texttt{NetworkX} \cite{networkx} to identify sets of $k$-regular graphs for even and odd $k$, as examples of graphs where the even and odd degree symmetries hold.  We then apply the symmetries of Table \ref{symmetrytable} to the angles for these graphs at $p=1$, giving the resulting degenerate sets of angles in Figure~\ref{fig:symmetry}. Here similarly colored blocks indicate regions where symmetry-related angles are found.  The original angles from the dataset have $\gamma_1 \in \Gamma=[-\pi/2,0]$ and $\beta_1 \in \mathrm{B}=[-\pi/4,0]$, where there is a set of optimized angles for almost all graphs \cite{lotshaw2021empirical}.  For $k$-regular graphs with $k$ even, these can be mapped to $\gamma_1 \to \gamma_1 \pm \pi$, giving the pairs of angles in Figure~\ref{fig:symmetry}(a).  A separate mapping under time-reversal symmetry  gives the pairs of angles in Figure~\ref{fig:symmetry}(b) shown for the same graphs; combining the even and time-reversal symmetries gives the sets of four angles in Figure~\ref{fig:symmetry}(c).  These can be contrasted with angles generated from $k$-regular graphs with $k$ odd, where there is joint symmetry in $\gamma_1$ and $\beta_1$  in Table \ref{symmetrytable}; combined with the time-reversal symmetry, this gives the sets of four angles in Figure~\ref{fig:symmetry}(d). The graphs with even and odd vertex degrees share two common sets of similar optimized angles when $|\gamma_1| \leq \pi/2$, while the other two sets of angles are in distinct regions of parameter space.  These confirm the importance of mapping optimized angles to similar intervals to reveal patterns in them, as highlighted using the example of differences between graphs with even and odd vertex degrees in~\cite{lotshaw2021empirical}.

\begin{figure}
    \centering
    \includegraphics[width=\linewidth]{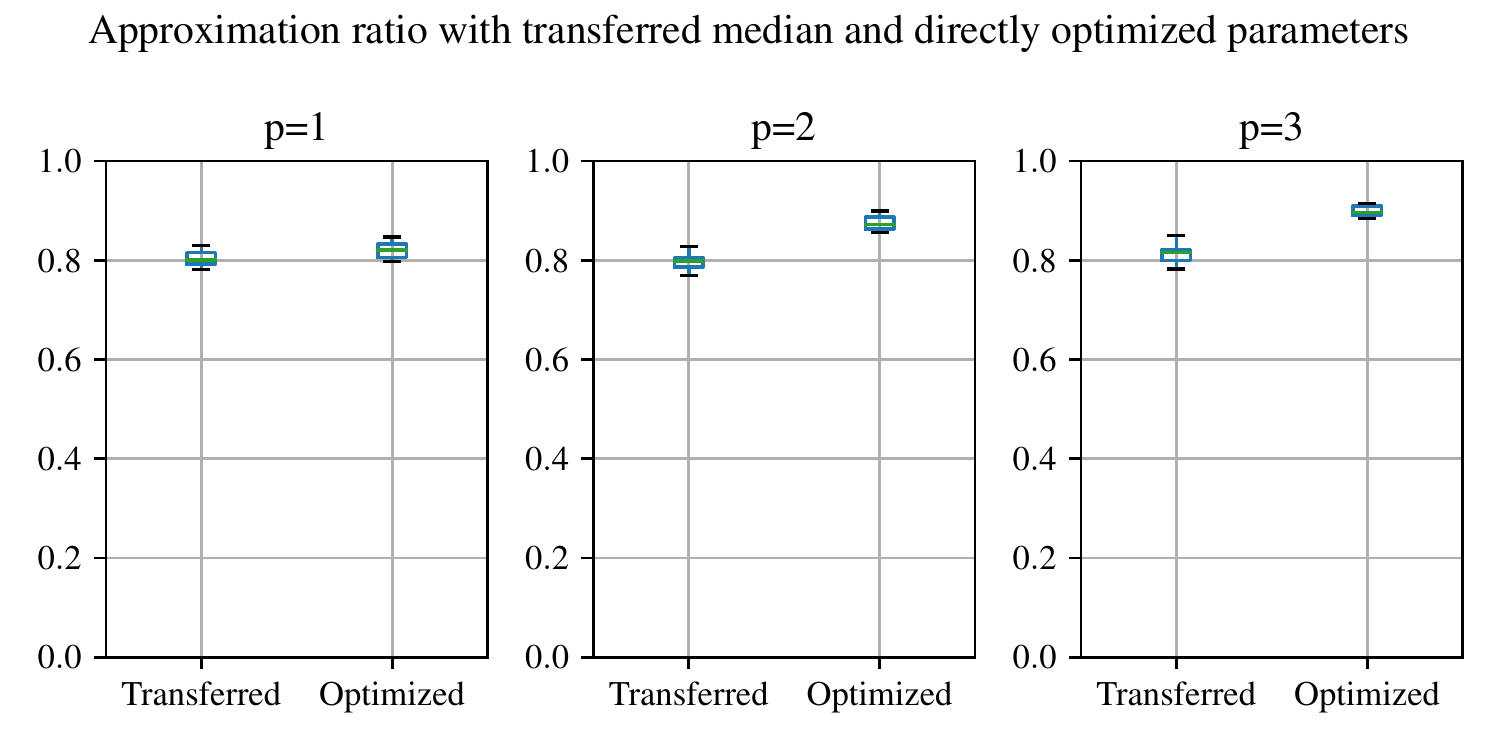}
    \caption{Approximation ratio obtained by the QAOA with a given $p$ using median angles over all non-isomorphic graphs with fewer than nine nodes (``Transferred'') and directly optimized angles (``Optimized'') for 10 random 20-node graphs. The box shows quartiles, the whiskers show maximum and minimum, and the green line shows the median approximation ratio.}
    \label{fig:transferability}
\end{figure}

\begin{figure}[ht]
    \centering
    \includegraphics[width=\linewidth]{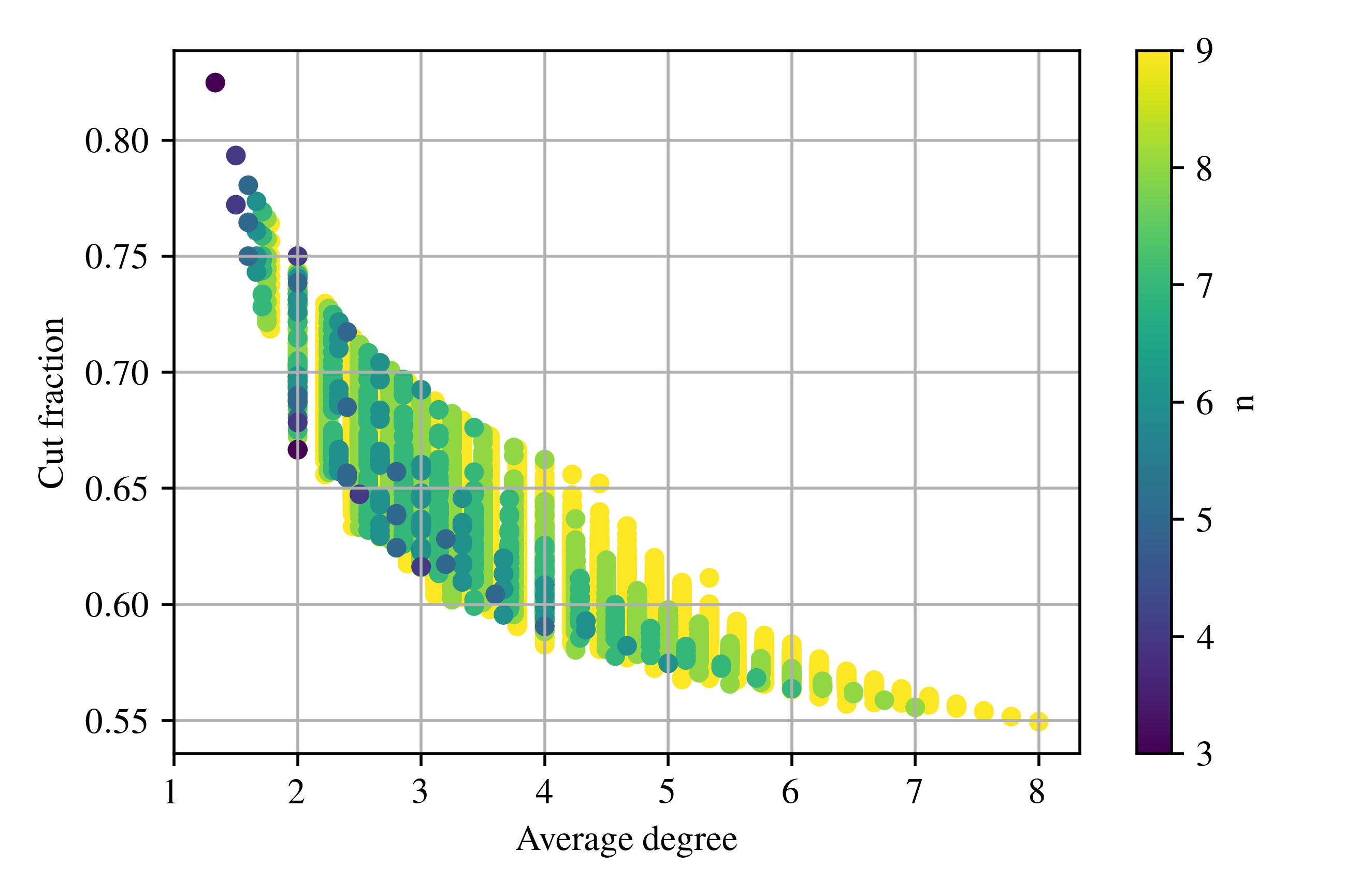}
    \caption{Plot of the cut fraction of the depth-1 QAOA on small graphs. The cut fraction decays with average degree, tending toward $1/2$.}
    \label{fig:cutfractionbyd}
\end{figure}

\subsection{Transferability of QAOA parameters} \label{sec:trasfer}

A well-documented consequence of concentration is that in the QAOA for MaxCut the optimal parameters are similar between similar graph instances; in other words, the optimal parameters from one instance can be \emph{transferred} to another one~\cite{brandao2018fixed-rs,Shaydulin2019MultistartDOI,Shaydulin2019EvaluatingDOI,khairy2019learning,2106.07531,2107.00677,lotshaw2021empirical}. More strongly, Lotshaw  et al.~\cite{lotshaw2021empirical} show that using one set of median (over the entire dataset at a given $n$ and $p$) angles gives approximation ratios that are identical or very close to the ratios  obtained from directly optimized angles. Approximation ratio here is defined as the ratio between the expected value of the cut from sampling the QAOA state and the true optimal cut. Since the dataset from \cite{lotshaw2021empirical} is included in \qaoakit{}, this finding can be easily reproduced and extended to new, previously unseen instances  of interest to the user. 

\begin{figure*}[t]
    \centering
    \includegraphics[width=\linewidth]{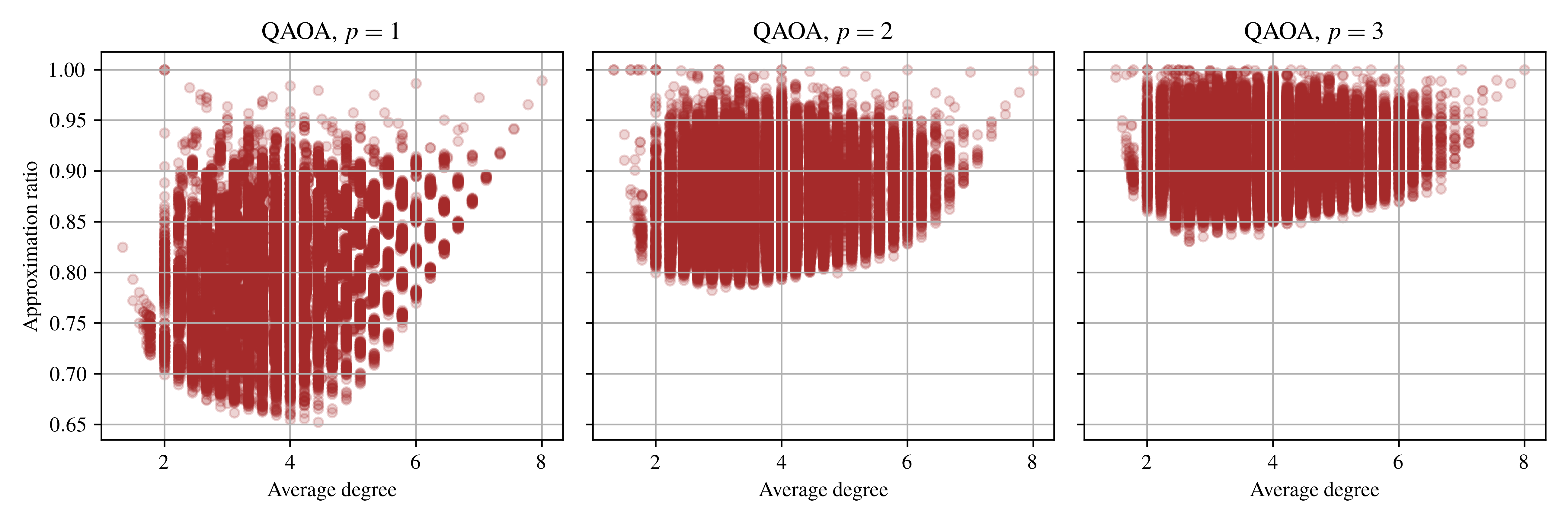}
    \caption{Plot of the approximation ratio of the QAOA on all graphs with up to 9 nodes. Each subplot is the result of the QAOA at a different depth. The worst-case approximation ratio improves with depth.}
    \label{fig:qaoabyp}
\end{figure*}

\begin{figure*}[t]
    \centering
    \includegraphics[width=\linewidth]{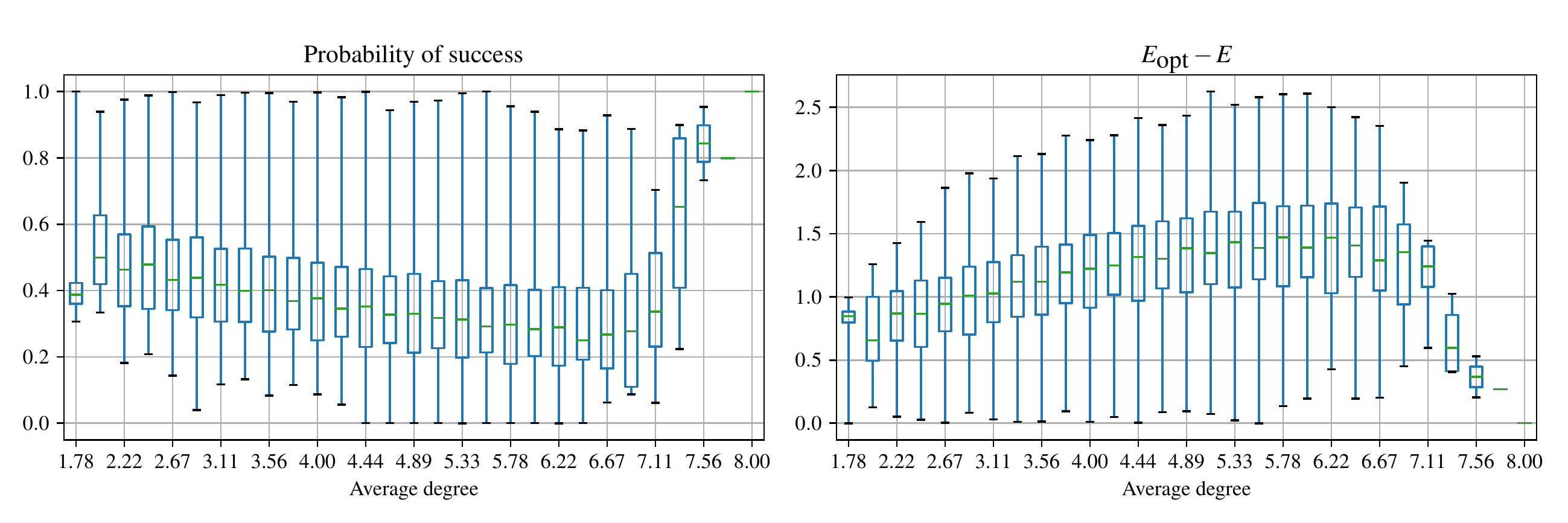}
    \caption{Plot of the success probability and the difference between expected cut produced by the QAOA $E$ and optimal cut value $E_{\text{opt}}$ for 9-node graphs with $p=3$. A negative (although weak) relationship between the success probability and average degree is clearly visible in the plots, supporting the findings of~\cite{Akshay2021}. When the average degree is very close to $n-1$, the QAOA has a much higher probability of success. This is likely do to the graph being nearly a clique and, consequently, the random solution (QAOA initial state) providing a higher approximation ratio. This behavior of the initial state ($p=0$) can also be observed in Figure~\ref{fig:symmetry_and_performance}.
    }
    \label{fig:biamonte}
\end{figure*}

We illustrate this feature in the example script \href{https://github.com/QAOAKit/QAOAKit/blob/master/examples/Transferability_to_unseen_instances.ipynb}{Transferability\_to\_unseen\_instances.ipynb}.
by picking $10$ random Erdos--Renyi model graphs with $20$ nodes and edge creation probability $0.5$. To illustrate the power of parameter transfer, we have found high-quality angles for these graphs using exhaustive optimization following the methodology of~\cite{lotshaw2021empirical}. The script computes the median angles over all 9-node graphs and plugs them into QAOA circuits for the $20$-node graphs, using the reference circuit implementations included in \qaoakit{}. Computing median parameters is possible because the angles are all in the same symmetry sector, as described in Section~\ref{sec:param_symmetries}. Median parameters provide a baseline performance,  which we can improve with further optimization. The directly optimized and transferred parameters are compared in Figure~\ref{fig:transferability}, which shows that  the median parameters provide an approximation ratio that is less than two percentage points away from the optimal for $p=1$.

Using median parameters from \qaoakit{} as the initial point can significantly reduce the cost of optimizing QAOA parameters  in practice. Advanced techniques such as training a generative model (for example, using kernel density estimation~\cite{khairy2019learning}) on the dataset can provide additional performance improvements.

\subsection{Exploring the performance of the QAOA} \label{sec:performance}
The cut fraction, which measures the number of edges cut over the total number of edges in the graph, can elucidate certain scaling behaviors that cannot be seen from just the approximation ratio. Since \qaoakit{} has the dataset of optimized QAOA parameters on every graph of at most $9$ nodes, we can compute the average degree and the expected cut fraction of each graph and analyze the relationship between these metrics and QAOA performance. Specifically, we can use this to compare the cut fraction of $p=1$ QAOA with the average degree.

In sparse random graphs, the value of the optimal cut scales as $1/2 + O(1/\sqrt{d})$ with the average degree $d$ of the graph~\cite{dembo2017extremal}, as $d$ gets large. This provides an upper bound on the cut fraction achieved by the QAOA, leading to the QAOA cut fraction also scaling roughly as  $1/2 + O(1/\sqrt{d})$. This prediction is consistent with ensemble analysis over all small graphs, shown in Figure~\ref{fig:cutfractionbyd}.

An alternative metric of performance is the approximation ratio, which measures the average number of edges cut by an approximate solution, over the number of edges cut by the exact solution. Figure~\ref{fig:qaoabyp} plots the approximation ratio of every graph of $\leq 9$ edges for $p=1,2,3$, as accessed by the \qaoakit{} dataset.
As $p$ increases, more graphs achieve the optimal approximation ratio, and the worst-case approximation ratio improves from about $0.65$ to about $0.825$ as $p$ goes from $1$ to $3$.

\begin{figure}[t]
    \centering %
    \includegraphics[width=0.76\linewidth, trim=0 0 0 1.3in]{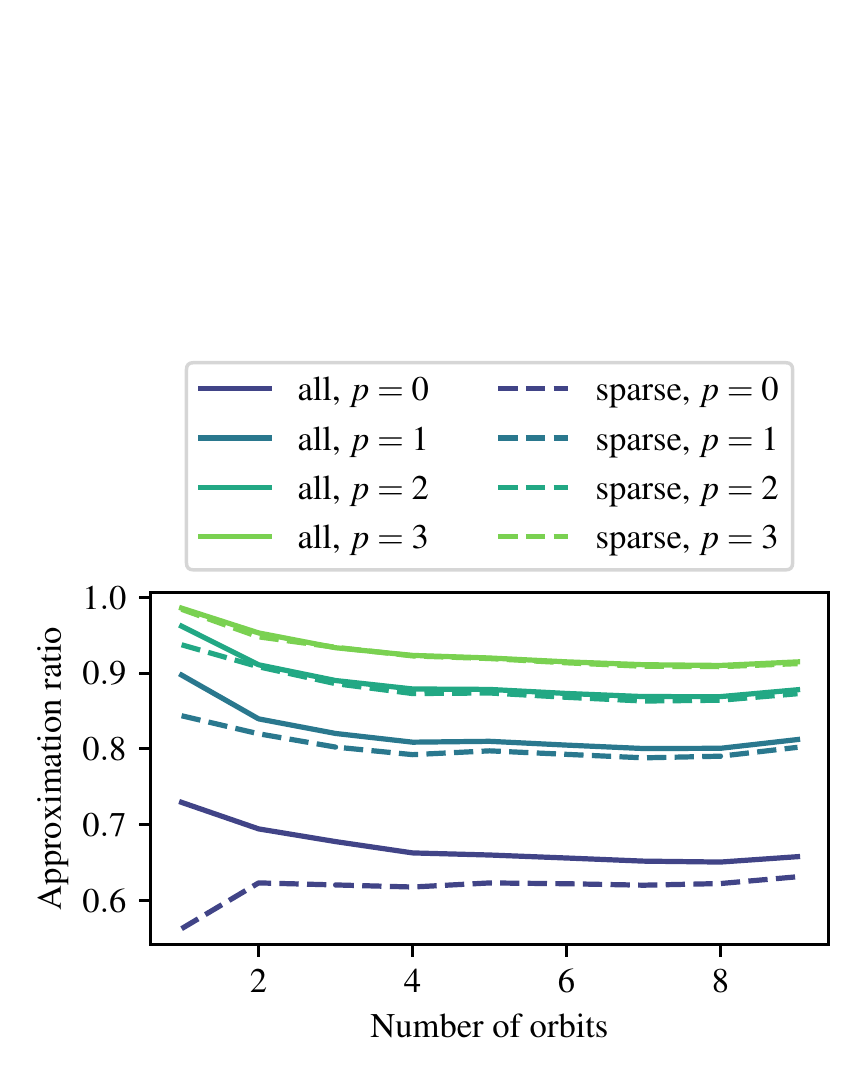}
    \caption{Relationship between QAOA approximation ratio and the amount of symmetry in the graph as measured by the number of vertex orbits. Solid lines present the mean approximation ratio over all 9-node graphs; dashed lines show only sparse graphs (average degree less than 4). $p=0$ line shows the approximation ratio achieved by the initial state. QAOA approximation ratio is highest for the most symmetric graphs (with the fewest vertex orbits).}
    \label{fig:symmetry_and_performance}
\end{figure}

Additionally, \qaoakit{} can easily be used to verify the claims about QAOA performance even if the dataset from a given paper is not directly available in \qaoakit{}. Akshay et al.~\cite{Akshay2021} claim that the performance of the QAOA exhibits rapid drop-off as the density of the problem (proportional to the average degree of the graph) increases. We can check whether this drop-off is exhibited for the regime available in the datasets included in \qaoakit{}. Specifically, we can consider all 9-node graphs and investigate the QAOA success probability and the difference between the optimal cut and the expected cut of the QAOA with $p=3$. Note that this is a lower QAOA depth than that considered in \cite{Akshay2021}. As can be observed in Figure~\ref{fig:biamonte}, in this regime a weak relationship is present, partially supporting the claims that the density of the problem affects QAOA performance.

The Jupyter notebook with this example is available at \href{https://github.com/QAOAKit/QAOAKit/blob/master/examples/performance.ipynb}{performance.ipynb}.

\subsection{Understanding the role of problem structure in QAOA performance} \label{sec:features}

Another application of \qaoakit{} may be identifying the properties of instances that are predictive of QAOA performance, which is an open problem in quantum optimization. For example, Shaydulin et al.~\cite{shaydulinsymm} suggest that symmetry may be one such property. %
In the example script \href{https://github.com/QAOAKit/QAOAKit/blob/master/examples/QAOA_symmetry_and_performance.ipynb}{QAOA\_symmetry\_and\_performance.ipynb}, we show how the datasets in \qaoakit{} can be used to study the relationship between symmetry and the approximation ratio achieved by the QAOA. We use the number of equivalence classes imposed by the group of automorphisms of a graph on its set of vertices (vertex orbits) as the symmetry measure. As can be observed from   Figure~\ref{fig:symmetry_and_performance}, which evaluates all nonisomorphic 9-node graphs for $p\in \{1,2,3\}$ (783,240 instances), the QAOA appears to perform better on average on more symmetric instances (with fewer orbits). This effect persists even on sparser graphs (average degree $\leq 4$) where the QAOA initial state approximation ratio does not depend on the symmetry of the instance. We note that the highly symmetric graphs ($\leq$ 3 vertex orbits) constitute only a small share of all graphs (535 out of 261,080 and 145 out of 108,247 if restricted to only graphs with average degree $\leq 4$). In that sense, the performance on highly symmetric graphs is not typical. However, no such relationship is observed for the 16-node 3-regular graphs present in the dataset, potentially due to the fact that there are very few symmetric 3-regular graphs. We encourage researchers to further scrutinize the data in \qaoakit{} to better understand the factors affecting QAOA performance.

\subsection{Comparing the QAOA with classical algorithms for MaxCut} \label{sec:classical}

\begin{figure}[t]
    \centering
    \includegraphics[width=\linewidth]{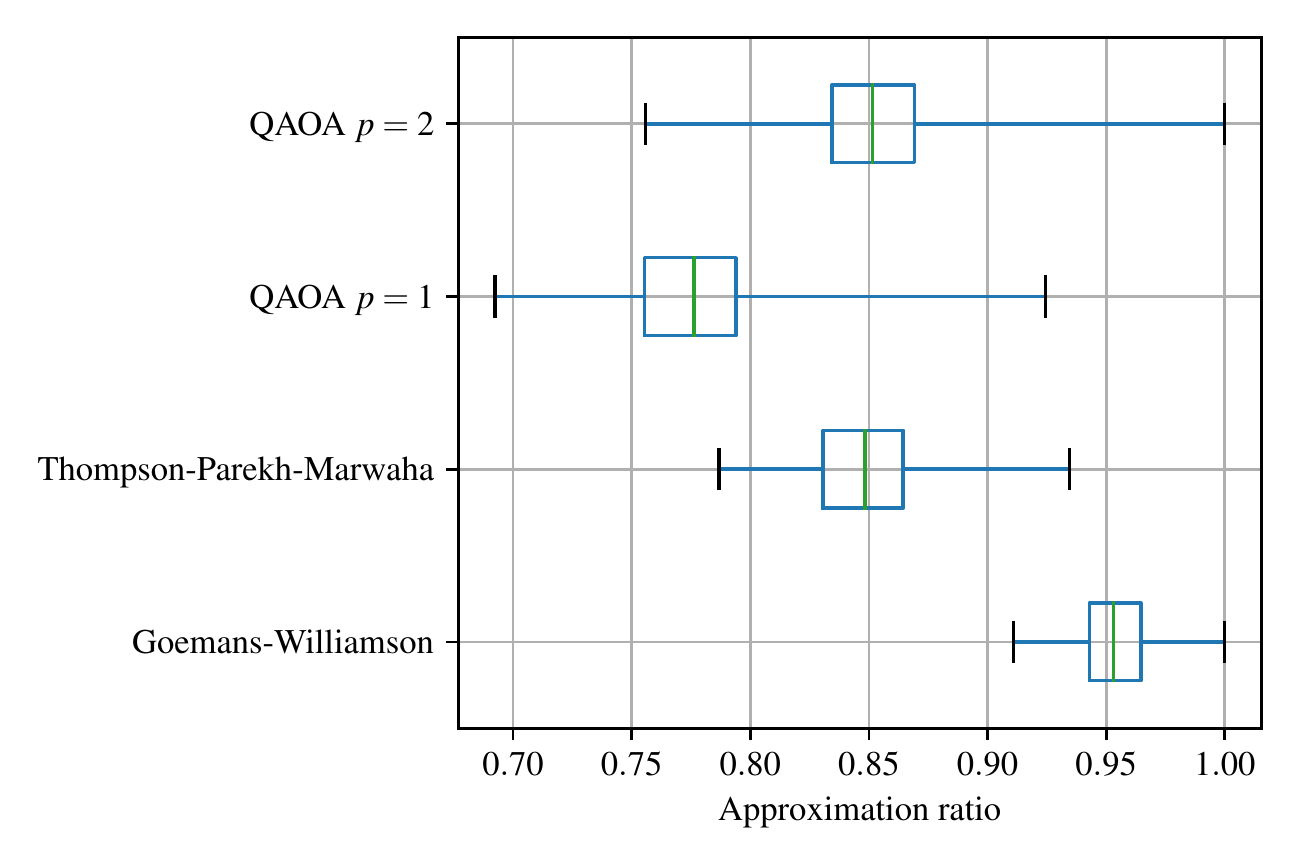}
    \caption{Approximation ratios of the QAOA and select classical algorithms for MaxCut on all non-isomorphic 3-regular graphs on 16 nodes. The box shows quartiles, the whiskers show maximum and minimum, and the green line shows the median approximation ratio.}
    \label{fig:classical_vs_quantum}
\end{figure}

The MaxCut problem has been studied extensively classically, with multiple classical algorithms available in the literature. \qaoakit{} includes an implementation of a recent explicit vector algorithm for high-girth MaxCut by Thompson et al.~\cite{Thompson2021explicit}. Additionally, we can include %
the classical Goemans--Williamson algorithm~\cite{goemans1995improved} in the comparison, with many implementations available.

\qaoakit{} contains approximation ratios obtained by the QAOA with $p\in\{1,2\}$ for all 3-regular graphs, making it easy to compare the performance of the QAOA with these algorithms. The example script \href{https://github.com/QAOAKit/QAOAKit/blob/master/examples/classical_algorithms_vs_qaoa.py}{classical\_algorithms\_vs\_qaoa.py} presents such a comparison. In Figure~\ref{fig:classical_vs_quantum} we observe that on average, the QAOA matches the performance of the algorithm in \cite{Thompson2021explicit} at $p=2$ on small 16-node graphs considered in this comparison. Note that the algorithm in that reference does not require solving a semi-definite programming problem, leading to lower runtime at the expense of lower approximation ratio as compared with the Goemans--Williamson algorithm. The value of $p$ required to achieve a given approximation ratio (e.g. ratio that matches that of~\cite{Thompson2021explicit}) is expected to grow with problem size. Therefore one should not expect that the QAOA with $p=2$ would match the performance of~\cite{Thompson2021explicit} at $p=2$ for all problem sizes.

\section{Discussion}

The rapid growth of literature on the Quantum Approximate Optimization Algorithm has created a need for standardized tools to reproduce and improve previously available results. This need is especially urgent when it comes to QAOA parameters, which are difficult to optimize~\cite{Shaydulin2019MultistartDOI} and in many settings must be optimized for a large number of instances. \qaoakit{} is a first step toward such a set of tools. 

We have provided specific examples demonstrating the versatility and effectiveness of \qaoakit{}.  First, we used \qaoakit{} to assemble large databases of pre-optimized QAOA parameters from previous studies and demonstrated how these can be used to analyze and reproduce claims related to parameter concentration, degeneracies, and symmetries.  Second, we showed a straightforward approach of using quantum simulation frameworks in conjunction with \qaoakit{} to generate new results exploring parameter transferability in QAOA, demonstrating the usefulness of \qaoakit{} as a tool for research. Third, we used \qaoakit{} to systematically analyze a variety of correlations between QAOA performance and graph features.  This demonstrated how the datasets compiled in \qaoakit{} can be used to quickly assess claims from the literature where data may not be readily available.  Finally, we included an example where QAOA results can be compared directly with leading classical solvers, to assess performance in ongoing research.  %

\qaoakit{} allows a researcher to skip the step of implementing and optimizing QAOA and proceed directly to study the phenomena of interest. For example, the parameters and circuits from \qaoakit{} have already been used to study error mitigation techniques for the QAOA~\cite{shaydulin2021error}.

We seed \qaoakit{} with datasets for unweighted MaxCut because it is the problem most commonly used to investigate the QAOA. However, other problems must be studied to better understand the practical utility of QAOA. We intend to extend \qaoakit{} accordingly and encourage researchers to contribute their own datasets. We hope that by making data conveniently accessible, \qaoakit{} will enable research that may unlock practical quantum advantage in optimization. %

\bibliographystyle{kunal_IEEEtran}
\bibliography{references}

\vspace{3em}

\small

\framebox{\parbox{\linewidth}{
The submitted manuscript has been created by UChicago Argonne, LLC, Operator of 
Argonne National Laboratory (``Argonne''). Argonne, a U.S.\ Department of 
Energy Office of Science laboratory, is operated under Contract No.\ 
DE-AC02-06CH11357. The manuscript is also authored by UT-Battelle, LLC under Contract No. DE-AC05-00OR22725 with the U.S. Department of Energy.
The U.S.\ Government retains for itself, and others acting on its behalf, a 
paid-up nonexclusive, irrevocable worldwide license in said article to 
reproduce, prepare derivative works, distribute copies to the public, and 
perform publicly and display publicly, by or on behalf of the Government.  The 
Department of Energy will provide public access to these results of federally 
sponsored research in accordance with the DOE Public Access Plan. 
http://energy.gov/downloads/doe-public-access-plan.}}

\end{document}